\documentclass[english,twocolumn,showpacs,nofootinbib,superscriptaddress,showpacs]{revtex4}
\usepackage{latexsym}
\usepackage{graphicx}
\usepackage{epsfig,subfig}
\usepackage{epstopdf}
\usepackage[active]{srcltx}
\usepackage{amsmath}
\usepackage{amssymb}
\usepackage{hyperref}
\usepackage{cleveref}
\usepackage{mathrsfs}

\begin{document}

\title{Diffraction in Time of Polymer Particles}
\author{A. Mart\'{\i}n-Ruiz}
\affiliation{Instituto de Ciencias Nucleares, Universidad Nacional Aut\'{o}noma de M\'{e}%
xico, 04510 M\'{e}xico, D.F., M\'{e}xico}
\email{alberto.martin@nucleares.unam.mx}

\begin{abstract}
We study the quantum dynamics of a suddenly released beam of particles using a background independent (polymer) quantization scheme. We show that, in the first order of approximation, the low-energy polymer distribution converges to the standard quantum-mechanical result in a clear fashion, but also arises an additional small polymer correction term. We find that the high-energy polymer behaviour becomes predominant at short distances and short times. Numerical results are also presented. We find that particles whose wave functions satisfy the polymer wave equation do not exhibit the diffraction in time phenomena. The implementation of a lower bound to the possible resolution of times into the time-energy Heisenberg uncertainty relation is briefly discussed.
\end{abstract}

\maketitle

\section{Introduction}

One of the main challenges in physics today is the search of a Quantum Theory of Gravity (QTG).  A major difficulty in the development of such theories is the lack of experimentally accessible phenomena that could shed light on the possible route to QTG.  Such quantum gravitational effects are expected to become relevant near the Planck scale, where spacetime  itself is assumed to be quantized. Compared to typical energy scales we are able to reach in our experiments, the Planck energy is extremely high, too high to hope to be able to test it directly, what makes it so difficult to test such effects.

Because the predictions of quantum mechanics have been verified experimentally to an extremely high degree of accuracy, a possible route to test quantum gravitational effects is through high-sensitivity measurements of well known quantum-mechanical phenomena, as any deviation from the standard theory is, at least in principle, experimentally testable. In this framework, with the best position measurements (being of the order $ \Delta x \sim 10 ^{-18} m$), at present sensitivities are still insufficient and quantum gravitational corrections remain unexplored. Despite this limitation, experimental verification of a common modification of the Heisenberg uncertainty relation that appears in a vast range of approaches to QTG, has been reported in \cite{nature}.

A background independent quantization scheme that arose in Loop Quantum Gravity (LQG), the so called Polymer Quantization (PQ), has been used to explore mathematical and physical implications of theories such as quantum gravity \cite{Hossain2, Chacon}. PQ may be viewed as a separate development in its own right, and is applicable to any classical theory whether or not it contains gravity. Its central feature is that the momentum operator $ p $ is not realized directly as in Schr\"{o}dinger quantum mechanics because of a built-in notion of discreteness, but arises indirectly through the translation operator $ U _{\lambda} \equiv e ^{i \frac{p \lambda}{\hbar}} $. Various approaches to QTG (such as LQG, String Theory and Non-Commutative Geometries) suggest the existence of a minimum measurable length, or a maximum observable momentum \cite{signature, discreteness, uncertainty}. In PQ a length scale is required for its construction, while for the gravitational case this is identified with Planck length, in the mechanical case it is just a free parameter.

In this paper we analyze the physical consequences of the PQ scheme in the dynamics of a well-known quantum transient phenomena: Diffraction in Time (DIT). DIT was discussed first by Moshinsky \cite{Moshinsky}. It is a phenomenon associated with the quantum dynamics of suddenly released quantum particles initially confined in a region of space \footnote{The original setting consisted of a sudden opening of a shutter to release a semi-infinite beam, and provided a quantum, temporal analogue of spatial Fresnel diffraction theory by a sharp edge.}. The hallmark of DIT consists of temporal and spatial oscillations of the quantum density profile \cite{Calderon}. The basic results of Moshinsky's shutter are reviewed in \cref{AppA}. In \cref{shutter} we consider the shutter problem in the framework of polymer quantum mechanics. In \cref{P-S transition} we show that the polymer result converges to the Moshinsky result in the appropriate limit. In \cref{WaveEquation} we demonstrate that no DIT effect arises when the wave function satisfies the polymer wave equation. Finally in \cref{discussion} we briefly discuss the implementation of a lower bound to the possible resolution of times into the time-energy Heisenberg uncertainty relation.

\section{The shutter problem in polymer quantum mechanics} \label{shutter}

The problem we shall discuss is the following: a monochromatic beam of polymer particles of mass $ m $ and momentum $ p >0 $, impinges on a totally absorber shutter located at the origin. If at $ t=0 $ the shutter is opened, what will be the transient polymer density profile at a distance $ x _{\mu} $ from the shutter?

To tackle this problem, we first restrict the dynamics to an equispaced lattice $ \gamma \left( \lambda \right) = \left\lbrace \lambda  n \vert n \in\mathbb{Z} \right\rbrace $. The spectrum of the position operator $ \left\lbrace x _{\mu} = \lambda \mu \right\rbrace $ consists of a countable selection of points from the real line, which is analogous to the graphs covering 3-manifolds in LQG. The polymer Hilbert space $ \textit{H} _{poly} $\footnote{The kinemathical Hilbert space can be written as $ \textit{H} _{poly} = L^{2} \left(  \mathbb{R} _{d}, d \mu _{d} \right) $, with $  d \mu _{d} $ corresponding Haar measure, and $ \mathbb{R} _{d} $ the real line endowed with the discrete topology}, consists of position wave functions which are nonzero only on the lattice, restricting the momentum wave functions to be periodic functions of period $ \frac{2 \pi \hbar}{\lambda} $ \cite{Hossain, Seahra}. Here $ \lambda $ is regarded as a fundamental length scale of the polymer theory.

[For a simpler analysis of the problem, hereafter we use the following dimensionless quantities for position, momentum, energy and time,
\begin{equation}
\mu \equiv \frac{x _{\mu}}{\lambda} \;\;  , \;\; \rho \equiv \frac{p \lambda}{\hbar} \;\; , \;\; \varepsilon \equiv \frac{m \lambda ^{2} E}{\hbar ^{2}} \;\; , \;\; \tau \equiv \frac{\hbar t}{m \lambda ^{2}} ,
\end{equation}
respectively. Also we use the notation $\psi _{\mu} \left( \tau \right) \equiv \psi \left( x _{\mu} , \tau \right) $ for the wave function in coordinate representation.]

For (non-relativistic) polymer particles, the wave function $ \psi _{\mu} \left( \tau \right) $ that represents the state of the beam of polymer particles for $\tau >0$, satisfies the time-dependent polymer Schr\"{o}dinger equation \cite{Corichi, Ashtekar}
\begin{equation}
2 i \frac{\partial}{\partial \tau} \psi _{\mu} \left( \tau \right) = 2 \psi _{\mu} \left( \tau \right)  - \psi _{\mu + 1} \left( \tau \right) - \psi _{\mu - 1} \left( \tau \right) , \label{POL SCHR EQ}
\end{equation}
and the initial wave function
\begin{equation}
\psi _{\mu} \left( \tau =0 \right) = e ^{i\rho \mu} \Theta \left( - \mu \right),  \label{IN WAVE FUNCT}
\end{equation}
where $ \Theta \left( y \right)  $ is the Heaviside step function, and the momentum $\rho \in \left[ 0, \pi \right) $ is a solution of the (free) polymer dispersion relation, $ \varepsilon \left( \rho \right) = \sin ^{2} \rho $, considering a fixed value of $ \varepsilon $. Note that the energy spectrum is bounded from above, and the bound depends on the length scale $ \lambda $.

For $ \tau >0 $ the shutter has been removed and the dynamics is free. By using the free polymer propagator \cite{Flores}, namely 
\begin{equation}
K _{\lambda} \left( \mu , \tau ; \nu , \tau _{0} \right) = i ^{\nu-\mu} J _{\nu-\mu} \left( \tau - \tau _{0} \right)  e^{-i \left( \tau - \tau _{0} \right) }, \label{FREE POLY PROP}
\end{equation}
the solution of (\ref{POL SCHR EQ}), subject to the initial condition (\ref{IN WAVE FUNCT}), is then
\begin{equation}
\psi _{\mu} \left( \tau \right) = e ^{-i \left( \tau - \rho \mu \right)}  \Phi _{\mu} \left( \rho, \tau \right), \label{PWF}
\end{equation}
where we have defined
\begin{equation}
\Phi _{\mu} \left( \rho, \tau \right) \equiv \sum _{\nu = - \infty} ^{- \mu} J _{\nu} \left( \tau \right) e ^{i \left(  \rho + \frac{\pi}{2} \right) \nu }, \label{parametric function}
\end{equation}
By using properties of Bessel functions, one can further check that $ \psi _{\mu} \left( \tau \right) $ satisfies (\ref{POL SCHR EQ}) and the initial condition (\ref{IN WAVE FUNCT}).

\begin{figure}[tbp]
\vspace{0.5 cm} 
\par
\begin{center}
\subfloat[\label{DT(t)rho/smallA}]{\includegraphics[scale=0.4, natwidth=640,natheight=480]{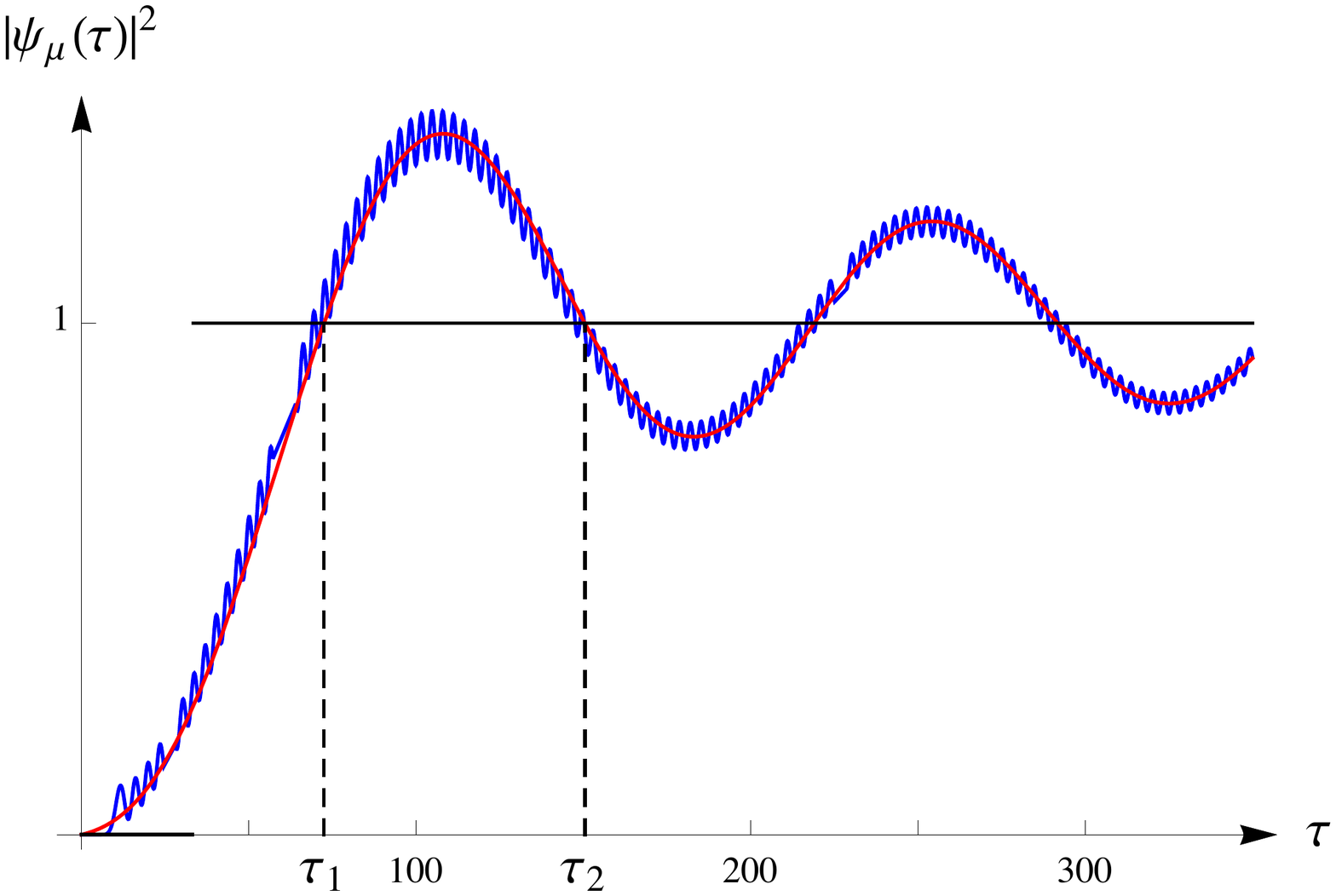}} \\[0pt]
\subfloat[\label{DT(t)rho/smallB}]{\includegraphics[scale=0.4,natwidth=640, natheight=480]{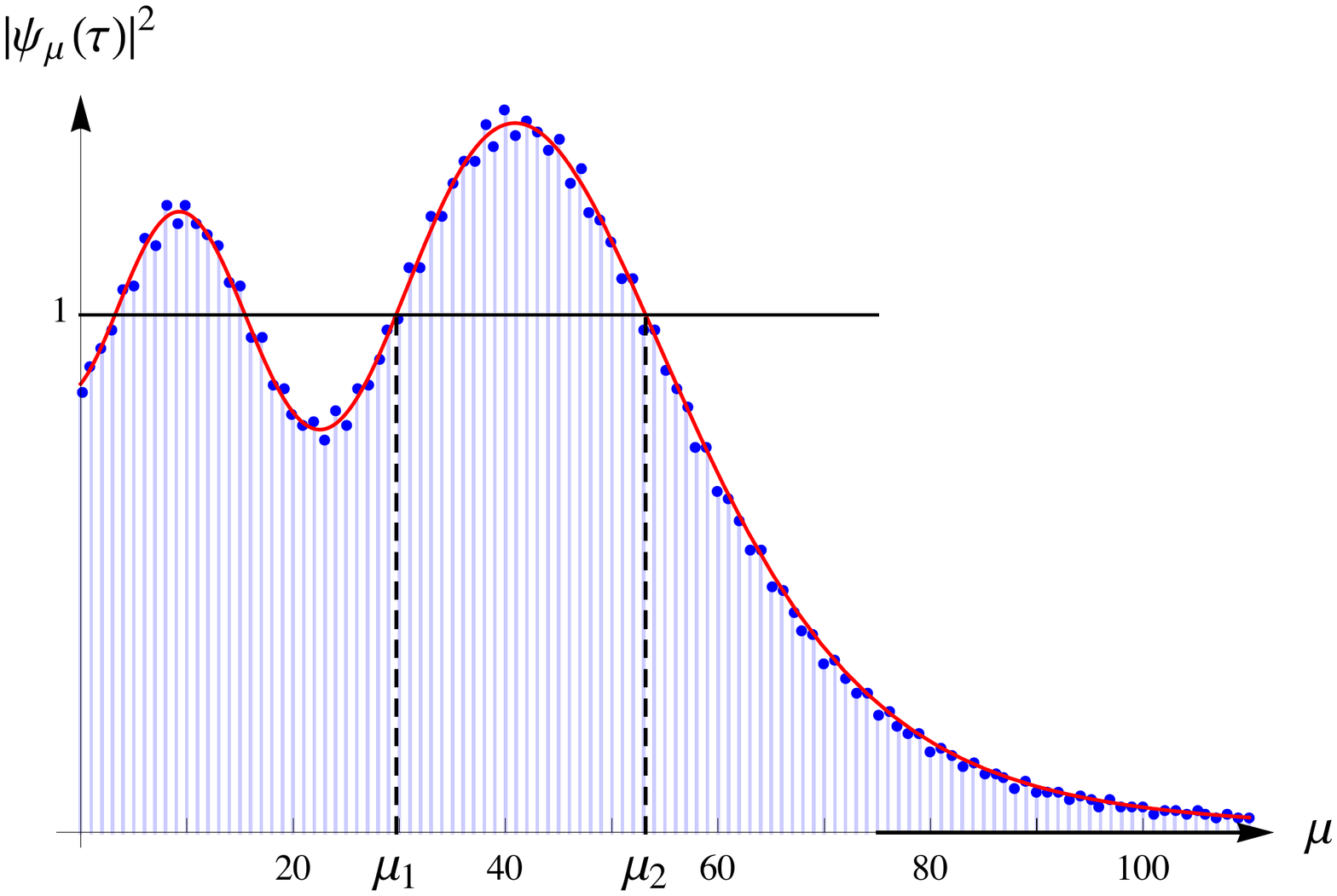}}
\end{center}
\caption{\small Here we plot the low energy ($ \rho = 0.3 $) polymer density profile as a function of time $ \tau $ at a fixed distance $ x _{10} $ [solid (blue) line in frame (a)], and as a function of $ \mu $ at a fixed time $ \tau = 250 $ [discrete (blue) plot in frame (b)]. In both graphics the solid (red) line corresponds to the Moshinsky distribution, and the solid (black) line corresponds to the classical result.} \label{DT(t)rho/small}
\end{figure}

\begin{figure}[tbp]
\vspace{0.5 cm}
\par
\begin{center}
\subfloat[\label{DT(t)rho/bigA}]{\includegraphics[scale=0.4, natwidth=640, natheight=480]{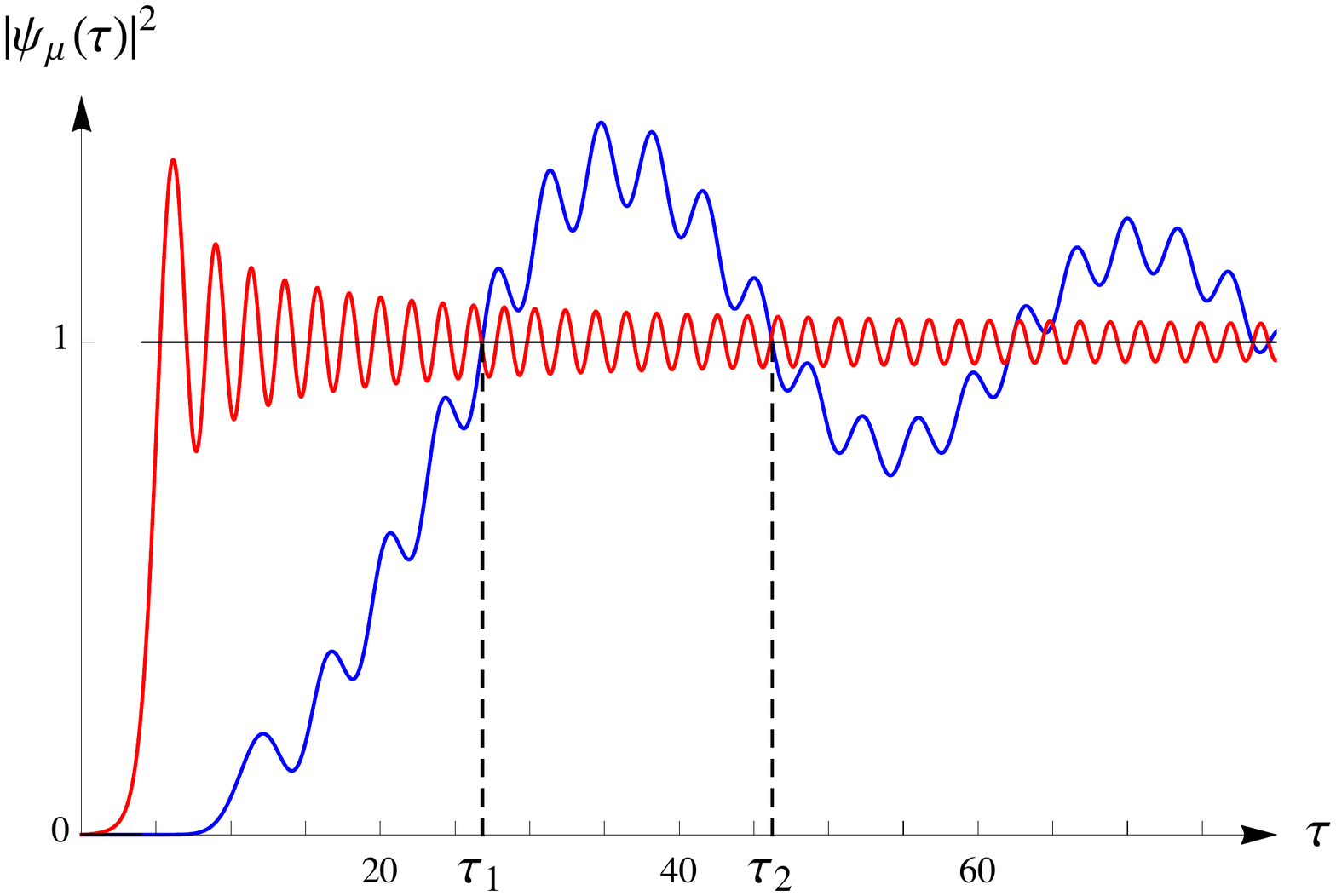}} \\[0pt]
\subfloat[\label{DT(x)rho/bigB}]{\includegraphics[scale=0.4, natwidth=640, natheight=480]{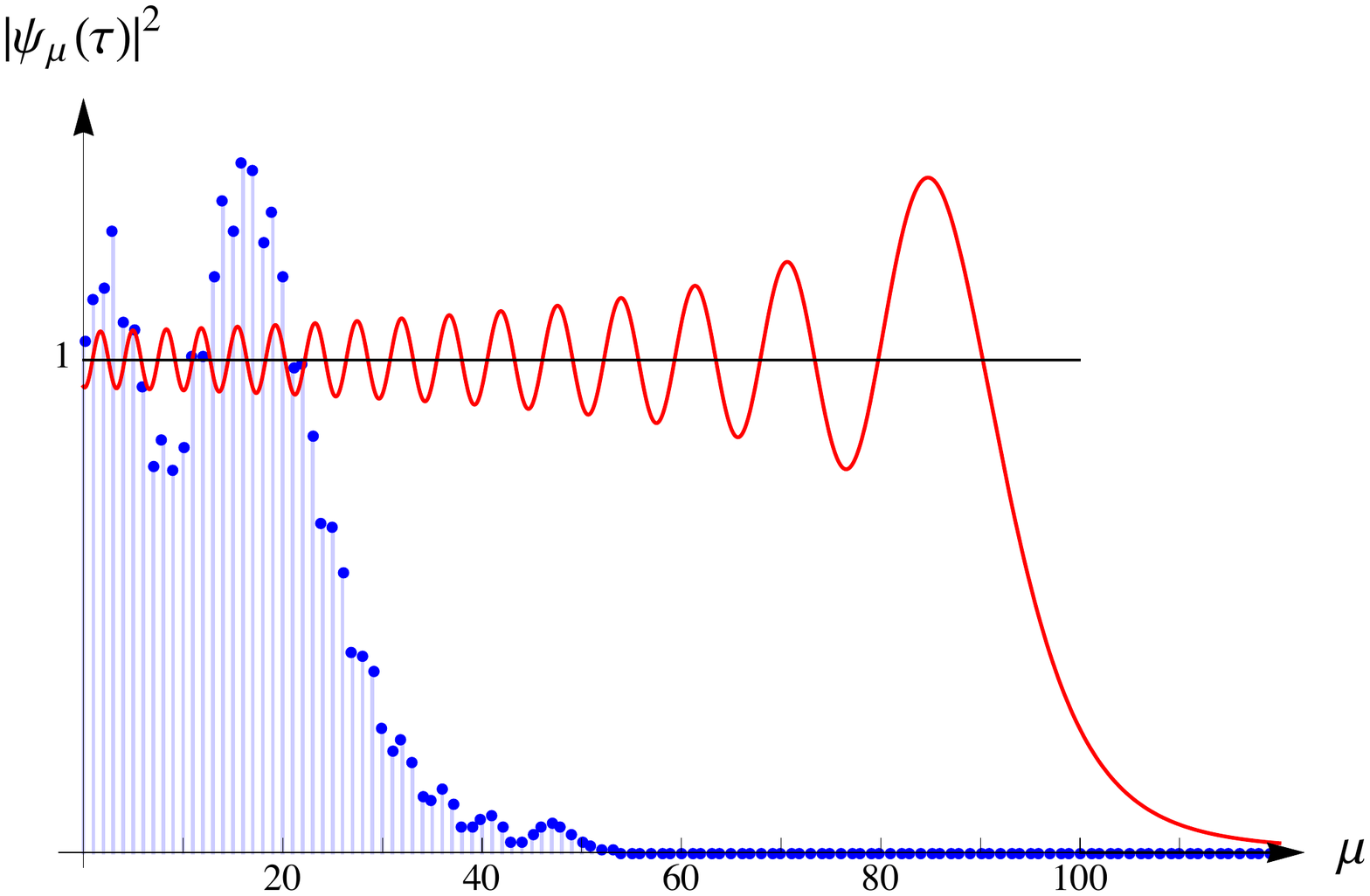}}
\end{center}
\caption{\small Here we plot the high energy ($ \rho = 2.5 $) Polymer density profile as a function of time $ \tau $ at a fixed distance $ x _{10} $ [solid (blue) line in frame (a)], and as a function of $ \mu $ at a fixed time $ \tau = 50 $ [discrete (blue) plot in frame (b)]. In both graphics the solid (red) line corresponds to the Moshinsky distribution, and the solid (black) line corresponds to the classical result.} \label{DT(t)rho/big}
\end{figure}

The corresponding polymer density profile,
\begin{equation}
\vert \psi _{\mu} \left( \tau \right)  \vert ^{2} =  \sum _{\nu , \alpha = - \infty} ^{- \mu} J _{\nu} \left( \tau \right) J _{\alpha} \left( \tau \right) \cos \left[ \left(  \rho + \frac{\pi}{2} \right) \left( \nu - \alpha \right) \right], \label{polymer profile}
\end{equation}
cannot be reduced to a simple form but they can be treated numerically. Plots of the polymer and the Moshinsky density profiles for both, the low and the high energy regimes, are presented in \cref{DT(t)rho/small,DT(t)rho/big}, respectively. At low energies ($ \rho \ll 1 $) the polymer and standard cases behave qualitatively in a similar manner. However, at high energies the polymer distribution exhibits some differences with respect to the standard case. Next we discuss the afore mentioned limiting cases.

In \cref{DT(t)rho/smallA} we plot the low energy polymer distribution as a function of time $ \tau $ at a fixed distance $ x _{\mu} $. We observe that the polymer result exhibits small oscillations superimposed on the Moshinsky result. This situation resembles the quantum-classical transition problem, where the classical distribution follows the spatial local average of the quantum probability density for large quantum numbers \cite{Bernal, Martin}. In this framework, \cref{DT(t)rho/smallA} suggests that the polymer and quantum-mechanical distributions approach each other in a locally time-averaged sense at low energies.

As in the standard case, a good measure of the width of the polymer diffraction effect in time, can be obtained from the difference $ \delta \tau $ between the first two times at which $ \vert \psi _{\mu} \left( \tau \right)  \vert ^{2} $ takes the classical value $ 1 $, i.e., $ \delta \tau = \tau _{2} - \tau _{1} $, as shown in \cref{DT(t)rho/smallA}. In this case such time difference can be estimated the same way as in the quantum-mechanical case (see \cref{AppA}) because at first order of approximation, the low-energy polymer wave function converges to the Moshinsky function evaluated on points in the lattice, i.e. $ \psi _{\mu} \left( \tau \right) \sim M \left( x _{\mu}, p, t \right) $ (see eq.(\ref{asymp MF})). By using the Cornu spiral one obtains $\delta \xi = 0.85 $, which leads to 
\begin{equation}
\delta \tau \simeq 0.85 \sqrt{\frac{\pi \mu}{\rho ^{3}}}, \label{tau width}
\end{equation}
for $ \rho \mu \gg  1 $. 

In this framework, the difference between both distributions, $ P \left( x _{\mu} , \tau \right) = \vert \psi _{\mu} \left( \tau \right)  \vert ^{2} - \vert M \left( x _{\mu}, p, t \right)  \vert ^{2} $, represents a good measure of the residual polymer behaviour at quantum level. It would be interesting that such deviations could be detected with high sensitivity experiments, however at low energies $ P \left( x _{\mu} , \tau \right) $ is enough small (of the order of $ \frac{1}{\sqrt{\tau}} $, see eq.(\ref{asymp MF})) to be detected in lab. For example, for the case considered in \cref{DT(t)rho/smallA} one obtains $ P _{ \texttt{max}} \simeq 0.0473868 \ll 1$, moreover taking $ \lambda $ in the order of the Planck length ($ l _{p} = 1.6 \times 10 ^{-35} m$), such deviation is extremely small ($P _{ \texttt{max}} \sim l _{p} $).

In \cref{DT(t)rho/smallB} we present the low energy polymer density profile as a function of $ \mu $ at a fixed time $ \tau $. As expected, we observe that the polymer result (discrete blue plot) resembles the standard result (continuous red line) as increasing $ \mu $, showing oscillations near the edge. The width of these oscillations can also be estimated the same way as in the standard case (see \cref{AppA}). With the help of the Cornu spiral one obtains
\begin{equation}
\delta \mu = 0.85 \sqrt{\pi \tau}. \label{mu width}
\end{equation}

The high energy polymer density profile as a function of time $ \tau $ at fixed distance $ x _{\mu} $ is presented in \cref{DT(t)rho/bigA}. Clearly, this case does not admit a simple analysis in terms of the Cornu spiral because the polymer distribution differs significatively with respect to the Moshinsky result. However we can construct a parametric like-spiral to perform a similar analysis. The standing point is that the polymer distribution can be expressed as $ \vert \operatorname{Re} \Phi \vert ^{2} + \vert \operatorname{Im} \Phi \vert ^{2} $, where $\operatorname{Im} $ and $ \operatorname{Re} $ stands for the imaginary and real parts of the function $ \Phi $ (\ref{parametric function}), respectively. Therefore we consider the curve that results from the parametric representation $ \left(  \operatorname{Re} \Phi , \operatorname{Im} \Phi \right) $.

\begin{figure}[tbp]
\vspace{0.5 cm} 
\par
\begin{center}
\includegraphics[scale=0.4, natwidth=640,natheight=480]{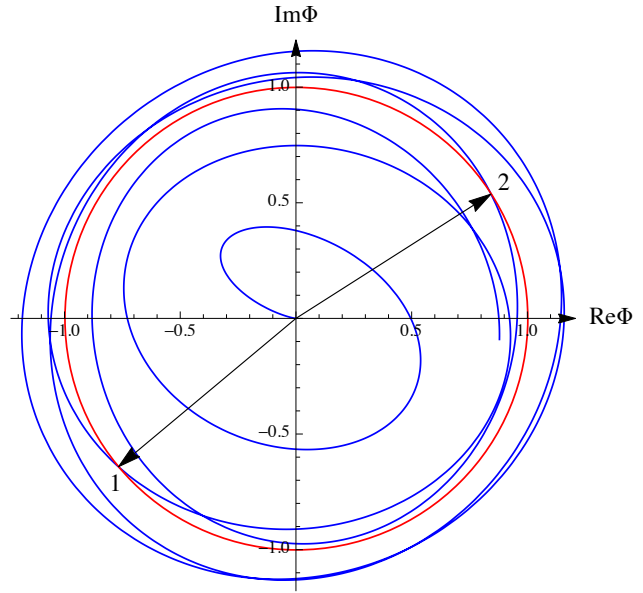}
\end{center}
\caption{\small Parametric like-spiral for $ \Phi _{10} \left( 2.5 , \tau \right) $} \label{spiral}
\end{figure}

In the like-spiral diagram of \cref{spiral}, the polymer probability density (\ref{polymer profile}) is the square of the radius vector from the origin $ \left( 0,0 \right) $ to the point on the spiral whose distance from the origin, along the curve, is $ \tau $. For the case considered in \cref{DT(t)rho/bigA}, when $\tau$ goes from $0$ to the classical time of flight $ \tau _{cl} = \small \frac{\mu}{\rho} $, the polymer distribution increases very slowly from $0$ to $ 4.71 \times 10 ^{-8} $. In other words, the classical time $\tau _{cl} $ is very small for detecting time-diffracted polymer particles.

The times at which the polymer distribution intersects the classical value $1$, correspond to the values of $ \tau $ obtained from the intersection of the like-spiral diagram with the circle of radius $1$ and center $ \left( 0,0 \right) $ in \cref{spiral}. The values of $\tau _{2} $ and $ \tau _{1} $ are the lenghts along the like-spiral from the origin to the points $1$ and $2$ in \cref{spiral}, so that we have $ \delta \tau \simeq 19 $, in agreement with the presented in \cref{DT(t)rho/bigA}. Of course, as increasing the energy the time width $ \delta \tau $ increases, but also the first time $ \tau _{1} $ at which the polymer distribution takes the classical value. Therefore high energy polymer particles also exhibit the diffraction effect in time, but with the characteristic times ($ \tau _{1}$ and $\delta \tau $) increased. In the limiting case $ \rho \rightarrow \pi $ the time $ \tau _{1} $ tends to infinite, and then the polymer particles (with the maximum possible energy) do not exhibit the diffraction effect in time. In the like-spiral diagram of \cref{spiral}, this case looks as a dense spiral completely contained in the unit circle.

In \cref{DT(x)rho/bigB} we plot the high energy polymer distribution as a function of $ \mu $ at a fixed time $ \tau $. We observe that the polymer distribution decreases abruptly before the particles reach the edge, and that the $ \mu _{2} $ value is smaller than for low energy polymer particles. Physically our results imply that the high energy polymer effects become important at short distances (\cref{DT(x)rho/bigB}) and short times (\cref{DT(t)rho/bigA}), as expected.

\section{The Polymer-Schr\"{o}dinger transition} \label{P-S transition}

It has been argued that if the lattice spacing $ \lambda $ is taken to be sufficiently small, the polymer formulation should reduce to the Schr\"{o}dinger representation \cite{Ashtekar}. This is a delicate issue because $ \lambda $ is regarded as a nonzero fundamental length scale of the polymer theory, and it cannot be removed when working in $ \textit{H} _{poly} $, no matter how small is $ \lambda $. This is analogous to the quantum-classical transition problem through $ \hbar \rightarrow 0 $ limit, because $ \hbar $ is a nonzero fundamental constant of the quantum theory \cite{Bernal, Martin}.

To address the Polymer-Schr\"{o}dinger transition, we consider the low energy regime of the polymer theory ($ \rho \ll 1 $), that one expects to be the domain of validity of the Schr\"{o}dinger theory. Clearly, the standard energy spectrum is recovered in the $ \rho \ll 1 $ limit, 
\begin{equation}
E \left( p \right) \approx \frac{p ^{2} }{2m} \ll \frac{2 \hbar ^{2}}{m \lambda ^{2}},
\end{equation} 
but it remains bounded from above due to the nonzero $ \lambda $. In terms of the de Broglie wavelength $ \lambda _{DB} = \frac{2 \pi \hbar}{p} $, this limit can be expressed as $ \lambda \ll \lambda _{DB} $, i.e. the fundamental length is very small compared with the characteristic quantum-mechanical length. Taking $ \lambda $ in the order of the Planck length, the typical diffraction experiments of electrons and neutrons, for which $ \lambda _{DB} \sim 10 ^{-10} m $, strongly satisfy the required condition \cite{Flores}.

Physically the $ \lambda \ll \lambda _{DB} $ limit implies that we should take the number of points between two arbitrary points very large, but keeping fixed its distance. Consequently for the distance between the shutter and the observer we must consider $ \mu \gg 1 $, and then the asymptotic behaviour of Bessel functions for large indices in \cref{PWF} is required. On the other hand, the time needed for the particle to move from $x _{\nu} $ to $x _{\mu}$ with momentum $ p $ is $ \tau = \frac{\vert \mu -\nu \vert }{\rho} $. Then by keeping fixed the distance, the $ \rho \ll 1$ limit implies that $\tau \gg 1$, and therefore the asymptotic behaviour of (\ref{PWF}) for large values of $ \tau $ is also required. Note that the argument of the Bessel function grows faster than its order, so that the $ \tau \gg 1 $ limit will dominate the transition \cite{Flores}.

The asymptotic expansion of the Bessel functions for large arguments is well known \cite{Watson}. It can be written as
\begin{equation}
J _{\nu} \left( \tau \right) \sim \frac{e ^{i \left[ \tau -\frac{\pi}{2} \left( \nu + \frac{1}{2} \right)  \right] }}{\sqrt{2 \pi \tau}} \left\lbrace R _{\nu} \left( \tau \right)   + R _{\nu} ^{\ast}\left( \tau \right)  e ^{-2i \left[ \tau -\frac{\pi}{2} \left( \nu + \frac{1}{2} \right)  \right] } \right\rbrace , \label{Asymp Bessel}
\end{equation}
where $ R _{\nu} \left( \tau \right) \equiv P _{\nu} \left( \tau \right) + i Q _{\nu} \left( \tau \right) $, with
\begin{eqnarray}
P _{\nu} \left( \tau \right) &=& \sum _{n=0} ^{\infty} \frac{\left( -1 \right) ^{n} }{\left( 2n \right)! \left( 2z \right) ^{2n}} \frac{ \Gamma \left( \nu + 2n + \frac{1}{2} \right) }{\Gamma \left( \nu - 2n + \frac{1}{2}\right)}, \\ \nonumber Q _{\nu} \left( \tau \right) &=& \sum _{n=0} ^{\infty} \frac{\left( -1 \right) ^{n} }{\left( 2n +1 \right)! \left( 2z \right) ^{2n + 1}} \frac{ \Gamma \left( \nu + (2n+1) + \frac{1}{2}\right) }{\Gamma \left( \nu - (2n+1) + \frac{1}{2}\right)} . \label{aux funct}
\end{eqnarray}

We observe that the expression (\ref{Asymp Bessel}) depends on the order of the Bessel function through both, the exponentials and the Gamma functions in (\ref{aux funct}). Fortunately only the second one becomes important for large indices $ \nu \gg 1$. Using the approximation $ \frac{ \Gamma \left( \nu + a + \frac{1}{2}\right) }{\Gamma \left( \nu - a + \frac{1}{2}\right)} \sim \nu ^{2a}$ for $ \nu \gg 1 $, the function $ R _{\nu} \left( \tau \right) $ becomes
\begin{equation}
R _{\nu} \left( \tau \right) \sim e ^{ i \frac{\nu ^{2}}{2 \tau}} .
\end{equation}
Finally the asymptotic behaviour of the Bessel function $ J _{\nu} \left( \tau \right)  $, for large arguments ($ \tau \gg 1 $) and large orders ($ \nu \gg 1 $), yields
\begin{equation}
J _{\nu} \left( \tau \right) e ^{-i \tau + i \frac{\pi \nu}{2}} \sim \frac{e ^{-i \frac{\pi}{4}}}{\sqrt{2 \pi \tau}} \left\lbrace e ^{i \frac{\nu ^{2}}{2 \tau}} + i \left( -1 \right) ^{\nu} e ^{ - i \frac{\nu ^{2}}{2 \tau}- 2i \tau} \right\rbrace . \label{Asymp Bessel 1}
\end{equation}
One can further check that in this regime the term proportional to $ e ^{ - 2 i \tau}  $ makes no contribution to the asymptotic behaviour because it produces a distributional expression, as pointed out in Ref.\cite{Flores}. 

Substituting (\ref{Asymp Bessel 1}) in (\ref{PWF}) we get a discrete version of the Moshinsky function,
\begin{equation}
\psi _{\mu} \left( \tau \right) \sim m \left( x _{\mu} , p, t \right) \equiv e ^{i \left( \rho \mu - \frac{\rho ^{2} \tau}{2} \right) } \frac{e ^{-i \frac{\pi}{4}}}{\sqrt{2 \pi \tau}} \sum _{\nu = - \infty} ^{\rho \tau -\mu} e ^{i \frac{\nu ^{2}}{2 \tau}}, 
\label{discrete M function}
\end{equation}
which satisfies analogous properties \cite{Calderon}, i.e.

\begin{itemize}
\item Under inversion of both $ x _{\mu} $ and $ p $, it satisfies
\begin{equation}
m \left( x _{\mu} , p , t \right) + m \left( - x _{\mu} , - p , t \right) = e ^{i \left( \rho \mu - \frac{\rho ^{2} \tau}{2} \right)} \vartheta _{3} \left( 0 , -2 \pi \tau \right) ,
\end{equation}
where $ \vartheta _{3} $ is the Jacobi's elliptic theta function \cite{Wittaker},
\item The asymptotic behaviour for $ \vert \rho \tau - \mu \vert \rightarrow \infty $ and $ 
\rho \tau \geq \mu $ is
\begin{equation}
m \left( x _{\mu} , p , t \right) \sim e ^{\frac{i}{\hbar} \left( p x _{\mu} - \frac{p ^{2}}{2m} t 
\right) },
\end{equation}
which is the standard quantum-mechanical result: a plane wave traveling to the right with momentum $ p $ and energy $ \frac{p ^{2}}{2m} $,
\item It satisfies the polymer Schr\"{o}dinger equation (\ref{POL SCHR EQ}).
\end{itemize}

To illuminate the correpondence between $ m \left( x _{\mu} , p, t \right) $ and the Moshinsky function (\ref{Moshinsky}) in a clear fashion, we first approximate the sum in (\ref{parametric function}) by using the Euler-Maclaurin formula,
\begin{eqnarray}
\Phi _{\mu} \left( \rho, \tau \right) \simeq  \int _{ - \infty} ^{- \mu} J _{\nu} \left( \tau \right) e ^{i \left(  \rho + \frac{\pi}{2} \right) \nu } d \nu + \frac{1}{2}  J _{- \mu} \left( \tau \right) e ^{- i \left(  \rho + \frac{\pi}{2} \right) \mu },
\end{eqnarray}
where we have used that $ J _{\nu} \left( \tau \right) \rightarrow 0 $ as $ \nu \rightarrow \infty $. Now, by the use of the asymptotic expansion of Bessel functions (\ref{Asymp Bessel 1}) we obtain
\begin{eqnarray}
m \left( x _{\mu} , p , t \right) &\cong & \frac{e ^{-i \frac{\pi}{4}}}{ \sqrt{2} } e^{ \frac{i}{\hbar} \left( p x _{\mu} - \frac{ p^{2}}{2m} t \right)} \left\lbrace  \left[ \frac{1}{2} + C\left( \xi \right)  \right] + i \left[  \frac{1}{2} + S \left( \xi \right) \right]  \right\rbrace  \\ \nonumber 
& & + \frac{1}{\sqrt{8 \pi \tau}} e ^{-i \left( \frac{\pi}{4} + \frac{\mu ^{2}}{2 \tau} \right)} \label{asymp MF}
\end{eqnarray}
where $ \xi \equiv \frac{1}{\sqrt{\pi \tau}} \left( \rho \tau - \mu \right) =\sqrt{\frac{m}{\pi \hbar t}} \left( \frac{pt}{m} -x _{\mu} \right) $. We recognize the first term in (\ref{asymp MF}) as the Moshinsky function (independent of $ \lambda $) evaluated on points in the lattice, i.e. $ M \left( x _{\mu} , p , t \right) $. The second term is small because it depends on $ \frac{1}{\sqrt{\tau}} \ll 1$ but it is nonzero. Of course this additional term represents the polymer residual behaviour at quantum level. Then at low energies we can neglect the correction term to analyze the polymer behaviour by using the Cornu spiral.

\section{POLYMER WAVE EQUATION} \label{WaveEquation}

Moshinsky also analyzed the shutter problem for different wave equations. The result was that no DIT phenomenon arises for the ordinary wave equation and the Klein-Gordon equation \cite{Moshinsky}. Of course this fact is intimately connected with the similarities between the Sommerfeld's diffraction theory and the time-dependent Schr\"{o}dinger equation.

In this section we shall consider the shutter problem, but assume that the state $\psi _{\mu} \left( \tau \right) $ satisfies the polymer wave equation
\begin{equation}
\frac{\partial ^{2} \psi _{\mu} \left( \tau \right) }{\partial \tau ^{2}} = \psi _{\mu +1} \left( \tau \right) + \psi _{\mu -1} \left( \tau \right) -2 \psi _{\mu} \left( \tau \right), \label{POL WAVR EQ}
\end{equation}
where $ \tau \equiv \frac{c t}{\lambda} $, and $c$ is the speed of light. Of course, we do not expect some resemblance between the solutions of (\ref{POL WAVR EQ}) and (\ref{POL SCHR EQ}). 

As usual, we will consider that initially $\psi _{\mu}$ and its time derivative are given by
\begin{equation}
 \psi _{\mu} \left( 0 \right) = F _{\mu} \;\ , \;\  \left(  \frac{\partial \psi _{\mu} \left( \tau \right) }{\partial \tau} \right) _{\tau = 0} = G _{\mu},
\end{equation}
and then the solution of (\ref{POL WAVR EQ}) can be obtained by using the Fourier integral theorem, i.e.
\begin{equation}
\psi _{\mu} \left( \tau \right) = \frac{1}{2 \pi} \int _{- \pi} ^{+ \pi} dk \left[ f \left( \kappa \right) \cos \left( \varepsilon _{\kappa} \tau \right) + g \left( \kappa \right) \frac{\sin \left( \varepsilon _{\kappa} t \right) }{\varepsilon _{\kappa}} \right] , \label{FOURIER INTEGRAL}
\end{equation}
where $f \left( \kappa \right)$ and $g \left( \kappa \right)$ are the Fourier transforms of $F _{\mu}$ and $G _{\mu} $, and $\varepsilon _{\kappa} = \sqrt{2 \left( 1 - \cos \kappa \right)} $. 

For $\tau <0$ the shutter was closed, and then we had on the left side of the shutter the simple (truncated) plane wave solution
\begin{equation}
\psi _{\mu} \left( \tau \right) = e ^{i \left( \rho \mu - \varepsilon _{\rho} \tau \right) } \Theta \left( -\mu \right). \label{PW WE}
\end{equation}
This implies that at $\tau =0$ we have:
\begin{equation}
F _{\mu} = e ^{i \rho \mu} \Theta \left( -\mu \right)   \;\ , \;\ G _{\mu} = -i \varepsilon _{\rho} e ^{i \rho \mu} \Theta \left( -\mu \right), \label{IN COND WE}
\end{equation}
while $F _{\mu} =G _{\mu} =0$ for $\mu > 0$. 

By Fourier transforming (\ref{IN COND WE}) we obtain that the initial conditions in the Fourier space become
\begin{eqnarray}
f \left( \kappa \right) &=& \pi \delta \left( \kappa -\rho \right) + i \cot \left( \kappa -\rho \right) , \\ \nonumber
g \left( \kappa \right) &=& - i \varepsilon _{\rho} f \left( \kappa \right). \label{IN COND WE MOM} 
\end{eqnarray}
These formulas are derived in \cref{IN COND FOUR SPACE}. We observe that in the low energy regime $f \left( \kappa \right)$ reduces to the $\delta _{+}$ function.

Substituting (\ref{IN COND WE MOM}) in (\ref{FOURIER INTEGRAL}) we get
\begin{eqnarray}
\psi _{\mu} \left( \tau \right) = \frac{1}{2} e ^{i \left( \rho \mu - \varepsilon _{\rho} \tau \right) } - \frac{1}{4 \pi i} \textbf{p.v.} \int _{- \pi} ^{+ \pi}  d \kappa \cot \left( \kappa - \rho \right) \times \\ \nonumber \left\lbrace \left( 1 + \frac{\varepsilon _{\rho}}{\varepsilon _{\kappa}} \right)  e ^{i \left( \kappa \mu - \varepsilon _{\kappa} \tau \right)  } + \left( 1 - \frac{\varepsilon _{\rho}}{\varepsilon _{\kappa}} \right) e ^{i \left( \kappa \mu + \varepsilon _{\kappa} \tau \right) } \right\rbrace \label{WAVE}
\end{eqnarray}

In this expression we intepret the integral in the sense of the Cauchy's principal value. In the $\kappa$-plane, the initial conditions can be obtained by closing the contour from above if $\mu > 0$ and from below if $ \mu < 0$. The integrand of (\ref{WAVE}) has two simple poles at $\kappa = \rho , \rho - \pi $ and an essential singularity at $\kappa = 0$, but the principal value of the  integral is convergent. In \cref{integral} we evaluate it. The result is 
\begin{eqnarray}
\psi _{\mu} \left( \tau \right) = \frac{1}{2} e ^{i \left( \rho \mu - \varepsilon _{\rho} \tau \right) } - \cot \rho \sin \frac{\rho}{2} + \frac{2}{\pi i } \sum _{\mu \neq \nu = - \infty} ^{+ \infty} \frac{J _{2 \nu} \left( 2 \tau \right) }{\mu + \nu} \times \\ \nonumber \Bigg[ 1 - {}_{2}F_{1} \left( 1, - \frac{\mu + \nu}{2}; 1 - \frac{\mu + \nu}{2}, e ^{2 i \rho} \right) \Theta \left( 2 - \mu - \nu \right) \\ \nonumber - {}_{2}F_{1} \left( 1, \frac{\mu + \nu}{2}; 1 + \frac{\mu + \nu}{2}, e ^{- 2 i \rho} \right) \Theta \left( 2 + \mu + \nu \right) \Bigg] , \label{SOL WAVE}
\end{eqnarray}
where $J _{n} $ are the Bessel functions of first kind and ${}_{2}F_{1}$ are the Gauss hypergeometric funtions in the unit circle. It is clear that, for $\rho$ and $\mu$ fixed, this function oscillates peridiocally in time. This behaviour has certainly no resemblance to the DIT effect obtained in \cref{shutter}.

\section{Discussion} \label{discussion}

The implications of the introduction of a nonzero fundamental length scale in quantum theory are quite profund. For example, the Heisenberg theorem, one of the cornerstones of quantum mechanics, states that the position $ x $ and momentum $ p $ of a particle cannot be simultaneously known with arbitrary precision, but the indeterminacies of a joint measurement are always bounded by $ \Delta x \Delta p \geq \frac{\hbar}{2}$. The implementation of such minimal length in quantum theory introduces a lower bound to the possible resolution of distances. It has therefore been suggested that the Heisenberg uncertainty relations should be modified to take into account the effects of spatial ``grainy" structure \cite{signature, discreteness, uncertainty}. The implementation of such ideas in polymer quantum mechanics is a difficult task because the momentum operator is not directly realized as in Schr\"{o}dinger quantum mechanics, furthermore the minimum length introduces an upper bound for momentum. In this framework, GUP theories predicts both a minimal observable length and a maximal momentum in the modified relation $\Delta x \Delta p \geq \frac{\hbar}{2} \left[ 1- 2 \hbar ^{-1} \Delta x _{ \texttt{min}} \Delta p + 4 \hbar ^{-2} \left( \Delta x _{ \texttt{min}}  \Delta p \right) ^{2} \right] $, with $ 2 \Delta p _{ \texttt{max}} \sim \hbar \left( \Delta x _{ \texttt{min}} \right) ^{-1} $.

It is well known that ordinary diffraction phenomena for beams of particles are closely associated with the position-momentum uncertainty relation, and the appearence of diffraction in time effects are closely connected with the time-energy uncertainty relation $ \Delta E  \Delta t \geq \frac{\hbar}{2} $. The possibility of closing the shutter after a time $ \Delta t $ have been used to form a pulse. For small values of $ \Delta t $ the time-energy uncertainty relation comes into play, broadening the energy distribution of the resulting pulse \cite{Moshinsky 2}. In the problem at hand, the lower bound to the possible resolution of distances introduces a minimal temporal window for the formation of a pulse of polymer particles. Therefore we suggest that also the time-energy uncertainty relation must be modified to implement the nonzero minimal uncertainty in time $ \Delta t _{ \texttt{min}} = \frac{m \lambda ^{2}}{\pi \hbar}$. The simplest generalized time-energy uncertainty relation which imples the appearance of a nonzero minimal uncertainty in time has de form $ \Delta E \Delta t \geq \frac{\hbar}{2} \left[ 1 + \alpha \left( \Delta E \right) ^{2} + \beta \right] $, where $ \alpha $ and $ \beta $ are positive and independent of $ \Delta E $ and $ \Delta t $. Possible extensions of this work include more complicated (time-dependent) shutter windows and the formation of polymer pulses. These could shed light on how we should modify the time-energy uncertainty relation to take into account the lower and upper bounds for the uncertainties in time and energy, respectively.

The implementation of both, the position-momentum and the time-energy modified uncertainty relations, could play an important role in other branches of physics. For example in the framework of local quantum field theories, the reason why we have ultraviolet divergencies is that in short time region $ \Delta t \rightarrow 0 $, the uncertainty with respect to energy increases indefinitely $ \Delta E \rightarrow \infty $, which in turn induces a large uncertainty in momentum $ \Delta p $.  The large uncertainty in momentum means that the particles states allowed in the short distance region $ \Delta x $ grows indefinitely as $ \left( \Delta E \right) ^{3} $ in $4$-dimensional space-time \cite{Tamiaki}. In these theories where there is no cutoff built-in, all those states are expected to contribute to amplitudes with equal strength and consequently lead to UV infinities. A theory which naturally provides the adequate modified uncertainty relations could shed light on the route for curing such UV divergences.

Finally, let us summarize our results. In \cref{shutter} we study the quantum dynamics of a suddenly released beam of polymer particles for both, the low and high energy regimes. Our numerical results show that in the quantum domain, the polymer distribution (as a function of time) exhibits small oscillations superimposed on the quantum-mechanical result (\cref{DT(t)rho/smallA}). Also the discrete spatial behaviour resembles the standard result for long distances, as shown in \cref{DT(t)rho/smallB}. In \cref{P-S transition} we show in an analytical clear fashion that in the first order of approximation the low energy polymer density profile converges to the Monshinsky distribution, but also emerges an additional polymer correction term, responsible of the small temporal and spatial deviations in \cref{DT(t)rho/small}. At high energies the diffraction effect in time also takes place, but only for long times (see \cref{DT(t)rho/bigA}). For polymer particles with the maximum possible energy the DIT is not longer present. Regarding to the spatial behaviour, the polymer distribution decreases abruptly before the particles reach the edge, as shown in \cref{DT(x)rho/bigB}. As expected, the polymer effects become important at short distances and short times. On the other hand we find that no diffraction effect in time arises when the wave functions satisfies the polymer wave equation.

\appendix

\section{The Moshinsky shutter} 
\label{AppA}

In the original Moshinsky's setup \cite{Moshinsky}, a quasi-monochromatic beam of non-relativistic quantum particles of momentum $ p $ is incident upon a perfectly absorbing shutter located at the origin and perpendicular to the beam. If the shutter is suddenly removed at $ t=0 $, what will be the transient density profile at a distance $ x $ from the shutter?

The initial wave function that we shall consider is
\begin{equation}
\psi \left( x, t=0 \right) = e^{i\frac{px}{\hbar}} \Theta \left( -x \right),  \label{initial WF}
\end{equation}
where $ \Theta \left( x \right)  $ is the Heaviside step function. Note that the state is not really monochromatic due the spatial truncation, besides that is clearly not normalized. Since for $ t > 0 $ the shutter has been removed, the dynamics is free and the time-evolved wave function is
\begin{equation}
\psi \left( x , t \right) = \int _{-\infty} ^{\infty} K \left( x,t;x^{\prime}, t^{\prime} =0 \right) 
\psi \left( x^{\prime}, t^{\prime} =0 \right) dx^{\prime}, \label{wave function}
\end{equation}
with the free propagator
\begin{equation}
K \left( x,t;x^{\prime}, t^{\prime} \right) = \sqrt{\frac{m}{2 \pi i \hbar \left( t - 
t^{\prime} \right) }} e^{i \frac{m \left( x - x^{\prime} \right) ^{2} }{2 \hbar \left( t - 
t^{\prime} \right) }}. \label{propagator}
\end{equation}

The solution of (\ref{wave function}) with the initial wave function (\ref{initial WF}) is known as Moshinsky functions,
\begin{equation}
M\left( x,p,t \right) = \frac{e ^{-i \frac{\pi}{4}}}{ \sqrt{2}  } e^{ \frac{i}{\hbar} \left( px - \frac{ p^{2}}{2m} t 
\right)} \left\lbrace \left[ \frac{1}{2} + C\left( \xi \right)  \right] + i \left[  \frac{1}{2} + S \left( 
\xi \right) \right]  \right\rbrace, \label{Moshinsky}
\end{equation}
where  $ C \left( \xi \right)  $ and $ S \left( \xi \right)  $ are the Fresnel integrals \cite{Gradshtein}, and $ \xi \equiv \sqrt{\frac{m}{\pi \hbar t}} \left( \frac{pt}{m} -x \right) $.

\begin{figure}[tbp]
\vspace{0.5 cm}
\par
\begin{center}
\includegraphics[scale=0.4, natwidth=640, natheight=480]{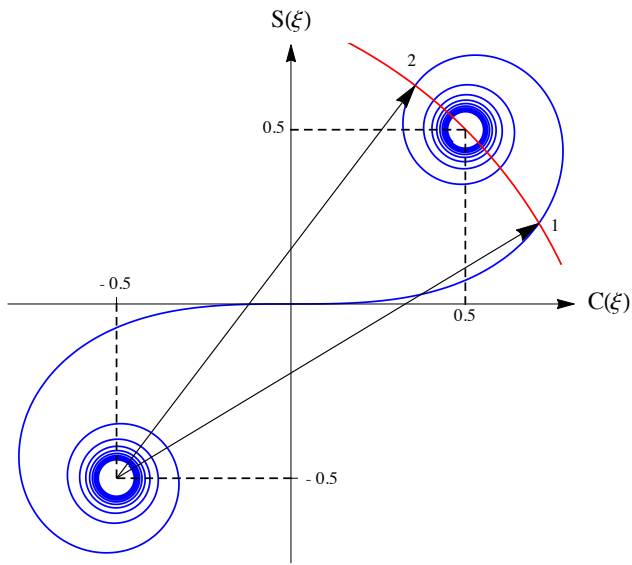}
\end{center}
\caption{\small Cornu spiral} \label{cornu}
\end{figure}

\begin{figure}[tbp]
\vspace{0.5 cm}
\par
\begin{center}
\subfloat[\label{DT-time}]{\includegraphics[scale=0.4, natwidth=640, natheight=480]{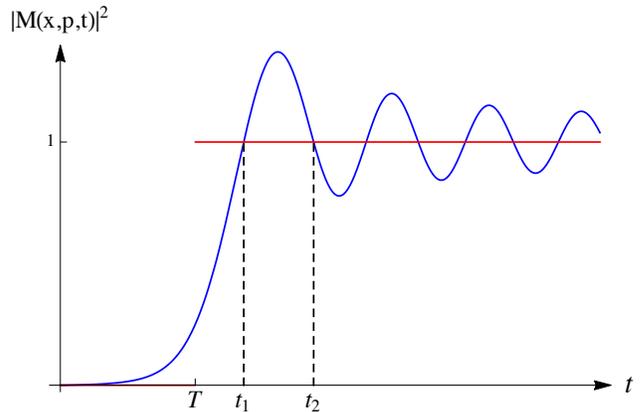}}
\subfloat[\label{DT-space}]{\includegraphics[scale=0.4, natwidth=640, natheight=480]{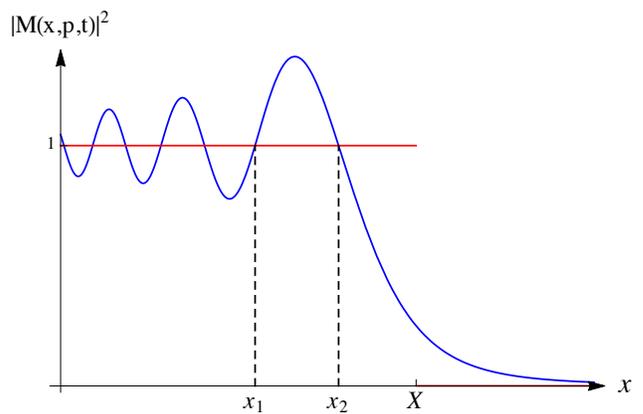}}
\end{center}
\caption{\small Here we plot the classical (red line) and quantum (blue line) density profiles as a function of time $ t $ at a fixed distance $ x $ [frame (a)] and as a function of position $ x $ at a fixed time $ t $ [frame (b)]} \label{DT}
\end{figure}

The ``Diffraction in Time'' term was introduced because the temporal behaviour of the quantum density profile (\cref{DT-time}),
\begin{equation}
\vert M\left( x,p,t \right) \vert ^{2} = \frac{1}{2} \left\lbrace \left[ \frac{1}{2} + C\left( \xi 
\right)  \right] ^{2} + \left[ \frac{1}{2} + S\left( \xi \right)  \right] ^{2} \right\rbrace , 
\label{quantum profile}
\end{equation}
admits a simple geometric interpretation in terms of Cornu spiral or clothoid, which is the curve that results from a parametric representation of the Fresnel integrals (\cref{cornu}). This is precisely the form of the intensity profile of a light beam diffracted by a semi-infinite plane \cite{Jackson}, which prompted the choice of the DIT term. The probability density is one-half of the square of the distance from the point $\left( - \small \frac{1}{2}, - \small \frac{1}{2} \right) $ to any other point of the Cornu spiral. In such representation the origin corresponds to the classical particle with momentum $ p $ released at time $ t=0 $ from the shutter position.

In \cref{DT-time} we present the classical and quantum density profiles at a fixed distance $ x $ as a function of time. The corresponding classical problem admits a trivial solution: the density profile vanishes if $ t $ is less than the time of flight $ T = \frac{mx}{p} $, but one if $ t \geq T $. On the other hand, with the help of the Cornu spiral we see that when $ t $ goes from $0$ to $T$, the quantum density profile increases monotonically from $0$ to $\small \frac{1}{4}$, while when $ t $ is larger than $ T $, then $ \vert M\left( x,k,t \right) \vert ^{2} $ behaves as a damped oscillation around the classical value, tending to this value when $ t \rightarrow \infty $.

The time width of this diffraction effect can be obtained from the difference $ \delta t $ between the first two times at which $ \vert M\left( x,p,t \right) \vert ^{2} $ takes de classical value, i.e., $ \delta t = t _{2} - t _{1} $, as shown in \cref{DT-time}. Such times correspond to the values of $ \xi $ obtained from the intersection of the Cornu spiral with the circle of radius $ \sqrt{2} $ and center $ \left( - \small \frac{1}{2} , - \small \frac{1}{2} \right) $. The difference $\delta \xi $ (that corresponds to $ \delta t $) can be estimated as the arc length along the Cornu spiral between the points $1$ and $2$ in \cref{cornu}, so that we have $\delta \xi = 0.85$. For $ px \gg \hbar $ the result is
\begin{equation}
\delta t \simeq 0.85 \sqrt{\frac{\pi \hbar }{px}} \; T. \label{time width}
\end{equation}

\Cref{DT-space} shows the classical and quantum density profiles at a fixed time $ t $ as a function of position. We see that an initial sharp-edge wave packet will move with the classical velocity showing rapid oscillations near the edge, located at $X = \frac{pt}{m}$. As before, the width of these oscillations can be estimated through the distance $ \delta x $ between the first two values of $ x $, starting from the edge, in which the probability density takes the classical value, i.e., $ \delta x = x _{2} - x _{1} $, as shown in \cref{DT-space}. Using the Cornu spiral we obtain $ \delta x = 0.85 \sqrt{\frac{\pi \hbar X}{p}} $. 

A more general type of initial state has been considered, 
\begin{equation}
\psi \left( x , t=0 \right) = e ^{i \frac{px}{\hbar}} + R e ^{-i \frac{px}{\hbar}},
\end{equation} 
with $ R = e ^{i \pi \alpha} $ corresponding to a shutter with reflectivity $ \vert R \vert ^{2} = 1 $. Under free evolution, $ \psi \left( x , t \right) = M \left( x , p , t \right) + R M \left( x , -p , t \right)$. For a complete review of the theory, results and experiments of quantum transients of single to few-body systems see Ref.\cite{Calderon}.

\section{Calculations of \cref{WaveEquation}}

\subsection{Derivation of $f \left( \kappa \right) $ and $g \left( \kappa \right) $} \label{IN COND FOUR SPACE}

Here we derive the initial conditions in momentum space by Fourier transforming (\ref{IN COND WE}), i.e.
\begin{equation}
f \left( \kappa \right) = \sum _{\mu = - \infty} ^{+ \infty} \Theta \left( - \mu \right) e ^{i \left( \rho - \kappa \right) \mu}. \label{IN WF MOM}
\end{equation}
In order to evaluate the sum, we consider the discretized derivative of the Heaviside step function,
\begin{equation}
 \frac{\partial \Theta \left( \mu \right)}{\partial \mu } \approx \frac{1}{2} \left[ \Theta \left( \mu +1 \right) - \Theta \left( \mu -1\right)\right].  \label{derivative theta}
\end{equation}
Note that this expression is equivalent to two delta functions at $\mu = \pm 1$. Then we find that 
\begin{equation}
\sum _{\mu = - \infty} ^{+ \infty}  \frac{\partial \Theta \left( \mu \right)}{\partial \mu } e ^{-i \zeta \mu} = \cos \zeta . \label{derivative theta 1}
\end{equation}
On the other hand, by substituting (\ref{derivative theta}) into (\ref{derivative theta 1}) and relabeling indices we find that
\begin{equation}
\sum _{\mu = - \infty} ^{+ \infty} \frac{\partial \Theta \left( \mu \right)}{\partial \mu } e ^{- i \zeta \mu} = i \sin \zeta \sum _{\mu = - \infty} ^{+ \infty} \Theta \left( \mu \right) e ^{-i \zeta \mu}. \label{derivative theta 2}
\end{equation}
By comparing (\ref{derivative theta 1}) and (\ref{derivative theta 2}) we obtain
\begin{equation}
\sum _{\mu = - \infty} ^{+ \infty} \Theta \left( \mu \right) e ^{-i \zeta \mu} = -i \cot \zeta + \pi \; \delta \left( \zeta \right) .  \label{delta +}
\end{equation}
The $\pi $ factor have been derived by using the property $\Theta \left( \mu \right) + \Theta \left( - \mu \right) = 1$. Finally with the help of (\ref{delta +}) we obtain the desired result, $f \left( \kappa \right) = i \cot \left( \kappa - \rho \right) + \pi \; \delta \left( \kappa - \rho \right) $. The second initial condition is proportional to $f \left( \kappa \right) $.

\subsection{The principal value} \label{integral}

Here we shall evaluate the integral (\ref{WAVE}). To this end, first we rewrite the integral as simple as possible. By using the well-known Jacobi-Anger expansion \cite{Gradshtein} we get
\begin{eqnarray}
\mathfrak{I} \equiv  \sum_{\nu = - \infty} ^{+ \infty}  J _{\nu} \left( 2 \tau \right) \int _{-\pi} ^{+\pi} d \kappa \cot \left( \kappa - \rho \right) e ^{i \kappa \varsigma } \times \\ \nonumber \left\lbrace \left[ 1 + (-1) ^{\nu} \right] - \left[ 1 - (-1) ^{\nu} \right] \sin \frac{\rho}{2} \csc \frac{\kappa}{2} \right\rbrace , \label{p.v. I}
\end{eqnarray}
where $J _{\nu} $ are the Bessel functions of first kind, and $\varsigma \equiv \mu + \frac{\nu}{2} $. We observe that, as in the standard case, the integrand has a simple pole at $\kappa = \rho $ and an essential singularity at $\kappa = 0 $, however in (\ref{p.v. I}) also emerges an additional simple pole at $\kappa = \rho - \pi $. 

As the lattice spacing is reduced, the solution of (\ref{p.v. I}) becomes a simple task. In this regime the limits of the integral open from $- \infty $ to $+ \infty$ and then the pole at $\rho - \pi $ should not be considered. Clearly the standard dispersion relation is recovered, $\varepsilon _{\kappa} \approx \kappa $, and the initial condition in Fourier space (\ref{IN COND WE MOM}) becomes the $\delta _{+}$ function, i.e. $f \left( \kappa \right)  \approx \pi \delta \left( \kappa - \rho \right) + i \left( \kappa - \rho \right) ^{-1} $. Under these conditions the principal value of $\mathfrak{I}$ can be computed directly by using the Cauchy's theorem. Moshinsky showed that no DIT phenomenon arises here.

\begin{figure}[tbp]
\vspace{0.5 cm}
\par
\begin{center}
\includegraphics[scale=0.4, natwidth=640, natheight=480]{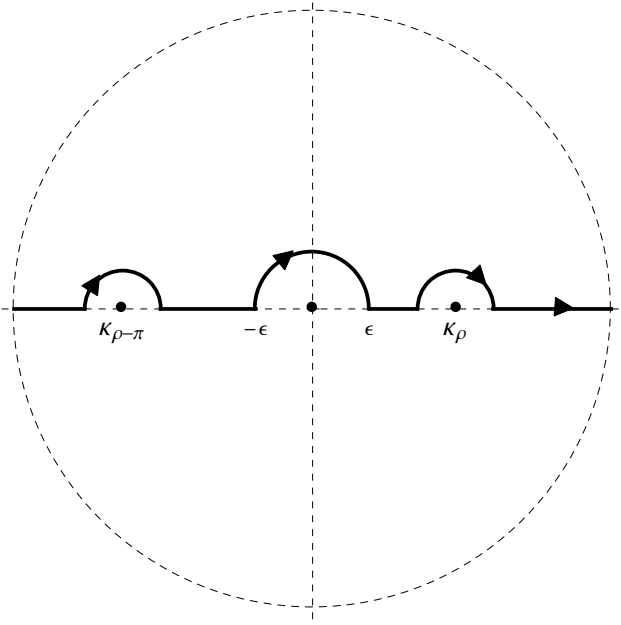}
\end{center}
\caption{\small $\kappa$-plane} \label{k-plane}
\end{figure}

Here we compute the principal value of $\mathfrak{I}$ by spliting the integral as follows
\begin{equation}
\textbf{p.v.} \; \mathfrak{I} = \lim _{\alpha , \epsilon , \beta \rightarrow 0 ^{+}}  \left(  \int _{-\pi} ^{\rho -\pi - \alpha}+ \int _{\rho - \pi +\alpha} ^{ - \epsilon} + \int _{+ \epsilon} ^{ \rho -\beta} + \int _{\rho + \beta} ^{\pi } \right) I d \kappa, \label{sep integral}
\end{equation}
where $I$ is the integrand in \ref{p.v. I}. See \cref{k-plane}.

We obtain straightforwardly that:
\begin{eqnarray}
& \mathfrak{I} _ {e} & \equiv \textbf{p.v.} \int _{-\pi } ^{+\pi} d \kappa \cot \left( \kappa - \rho \right) e ^{i \kappa \varsigma} = - \frac{4 }{\varsigma} \times \\ \nonumber  && \Big[ 1 -  {} _{2} F _{1} \left( 1 , - \frac{\varsigma}{2} ; 1 - \frac{\varsigma}{2}; e ^{2i \rho} \right) \Theta \left( 2 - \varsigma\right) \\ \nonumber  && -  {} _{2} F _{1} \left( 1 , \frac{\varsigma}{2} ; 1 + \frac{\varsigma}{2};  e ^{-2i \rho} \right) \Theta \left( 2 + \varsigma\right)\Big], \label{even int}
\end{eqnarray}
where $ {} _{2} F _{1} \left( a,b;c;z \right) $ are the hypergeometric Gauss functions in the unit circle $\vert z \vert = 1$. The Heaviside step functions in (\ref{even int}) ensure the convergence of the corresponding hypergeometric functions. In \cref{p.v. I} we can see that the term $\mathfrak{I} _ {e}$ is multiplied by $\left[ 1 + (-1) ^{\nu} \right]$, and therefore its product is restricted to even values of $\nu$.

On the other hand we also obtain that:
\begin{eqnarray}
\mathfrak{I} _ {o} \equiv \textbf{p.v.} \int _{-\pi } ^{+\pi} d \kappa \cot \left( \kappa - \rho \right) \csc \left( \frac{\kappa}{2} \right)  e ^{i \kappa \varsigma} = - 2 \pi i \cot \rho  \\ \nonumber - 2 \cos \left( \pi \varsigma \right) \cot \rho \Bigg[ e ^{i \pi \varsigma} \Big( B \left( -i;2 \varsigma , 0\right) + B \left( -i;1 +2 \varsigma , 0\right) \Big) \\ \nonumber - e ^{- i \pi \varsigma} \Big( B \left( i;2 \varsigma , 0\right) + B \left( i;1 +2 \varsigma , 0\right) \Big) + 4 i \tan \rho \times \\ \nonumber \left(  \frac{{}_{2} F _{1} \left( 1 , \frac{1 + 2 \varsigma}{4}; \frac{5 + 2 \varsigma}{4}; e ^{-2i \rho}\right) }{1 + 2 \varsigma} - \frac{{}_{2} F _{1} \left( 1 , \frac{3 + 2 \varsigma}{4}; \frac{7 + 2 \varsigma}{4}; e ^{-2i \rho}\right) }{3 + 2 \varsigma} \right)  \Bigg] , \label{odd int}
\end{eqnarray}
where $B \left( z ; a ,b \right) $ are the incomplete Beta functions in the unit circle. Here we have omited the Heaviside functions which ensures de convergenge of each term. In \ref{p.v. I} we observe that this term is multiplied by $\left[ 1 -  (-1) ^{\nu} \right]$, restricting its product to odd values of $\nu$. Note that under this condition $\cos \left(\pi \varsigma \right) = 0$, and therefore only the first term in (\ref{odd int}) contributes to the integral.

By substituting (\ref{even int}) and (\ref{odd int}) into (\ref{p.v. I}) we finally obtain that
\begin{eqnarray}
\textbf{p.v.} \mathfrak{I} &=& 4 \pi i \cot \rho \sin \frac{\rho}{2}  - 8 \sum_{\nu = - \infty} ^{+ \infty} \frac{ J _{2 \nu} \left( 2 \tau \right)}{\varsigma ^{\prime}} \times \\ \nonumber && \Bigg[ 1 - {} _{2} F _{1} \left( 1 , - \frac{\varsigma ^{\prime}}{2} ; 1 - \frac{\varsigma ^{\prime}}{2}; e ^{2i \rho} \right) \Theta \left( 2 - \varsigma ^{\prime} \right) \\ \nonumber && \;\;\;\  - {} _{2} F _{1} \left( 1 , \frac{\varsigma ^{\prime}}{2} ; 1 + \frac{\varsigma ^{\prime}}{2};  e ^{-2i \rho} \right) \Theta \left( 2 + \varsigma ^{\prime} \right) \Bigg]  
\end{eqnarray}
with $\varsigma ^{\prime} = \mu + \nu $. This result stablishes (\ref{SOL WAVE}).

\section*{Acknowledgements}
The author would like to thanks L. F. Urrutia, A. Frank, A. Carbajal, J. Bernal and E. Chan for useful discussions, comments and suggestions.


\begin{thebibliography}{99}

\bibitem{nature} I. Pikovski,  M. R. Vanner,	 M. Aspelmeyer, M. S. Kim and \v{C}. Brukner, Nature Physics \textbf{8}, 393 (2012).

\bibitem{Hossain2} G. M. Hossain, V. Husain, S. S. Seahra, Phys. Rev. D \textbf{81}, 024005 (2010).

\bibitem{Chacon} G. Chac\'{o}n-Acosta, E. Manrique, L. Dagdug and H. A. Morales-T\'{e}cotl, SIGMA \textbf{7}, 100 (2011).

\bibitem{signature} S. Hossenfelder, M. Bleicher, S. Hofmann, J. Ruppert, S. Scherer and H. St\"{o}cker, Physics Letters B \textbf{575}, 85 (2003).

\bibitem{discreteness} A. F. Ali, S. Das and E. C. Vagenas, Physics Letters B \textbf{678}, 497 (2009).

\bibitem{uncertainty} A. Kempf, G. Mangano and R. B. Mann, Phys.Rev. D \textbf{52}, 1108 (1995).

\bibitem{Moshinsky} M. Moshinsky, Phys. Rev. \textbf{88}, 625 (1952).

\bibitem{Calderon} A. del Campo, G. Garc\'{i}a-Calder\'{o}n and J.G. Muga, Physics Reports \textbf{476}, 1 (2009).

\bibitem{Hossain} G. M. Hossain, V. Husain, S. S. Seahra, Phys. Rev. D \textbf{82}, 124032 (2010).

\bibitem{Seahra} S. S. Seahra, I. A. Brown, G. M. Hossain and V. Husain, Journal of Cosmology and Astroparticle Physics \textbf{2012}, 041 (2012).

\bibitem{Corichi} A. Corichi, T. Vuka\v{s}inak and J. A. Zapata, Phys. Rev. D \textbf{76}, 044016 (2007).

\bibitem{Ashtekar} A. Ashtekar, S. Fairhurst and J. L. Willis, Classical and Quantum Gravity \textbf{20}, 1031 (2003).

\bibitem{Flores} E. Flores-Gonz\'{a}lez, H. A. Morales-T\'{e}cotl and J. D. Reyes, Annals of Physics \textbf{336}, 394 (2013).

\bibitem{Gradshtein} I. S. Gradshteyn and I. M. Ryzhik, \textit{Table of Integrals, Series, and Products}, Edited by A. Jeffrey and D. Zwil- linger, 4th Edition (Academic Press, New York, 1994).

\bibitem{Bernal} J. Bernal, A. Mart\'{i}n-Ruiz and J. Garc\'{i}a-Melgarejo, Journal of Modern Physics \textbf{4}, 108 (2013).

\bibitem{Martin} A. Mart\'{i}n-Ruiz, J. Bernal and A. Carbajal-Dom\'{i}nguez, Journal of Modern Physics \textbf{5}, 44 (2014).

\bibitem{Watson}  G. N. Watson, \textit{A Treatise on the Thery of Bessel Functions}, 2nd ed. (The University Press, 1994), p. 192

\bibitem{Wittaker} E. T. Whittaker and G. N. Watson, \textit{A Course of Modern Analysis}, 4th edition , (Cambridge University Press, 1927).

\bibitem{Moshinsky 2} M. Moshinsky, Am. J. Phys. \textbf{44}, 1037 (1976).

\bibitem{Tamiaki} Y. Tamiaki, Prog. Theor. Phys. \textbf{103}, 1081 (2000).

\bibitem{Jackson} J. D. Jackson, \textit{Classical Electrodynamics}, 3rd Edition (Wiley, 1998).

\end{thebibliography}
\end{document}